\documentclass[11pt,onecolumn]{IEEEtran}
\usepackage{subfigure}
\usepackage{epsfig,graphicx,psfrag}
\usepackage{amsfonts,amsmath,amssymb}

\newtheorem{theorem}{Theorem}
\newtheorem{lemma}{Lemma}
\newtheorem{corollary}{Corollary}

\newtheorem{definition}{Definition}
\newtheorem{note}{Note}
\newtheorem{property}{Property}

\newcommand{\rf}{\right}
\newcommand{\lf}{\left}
\renewcommand{\epsilon}{\varepsilon}
\newcommand{\Reals}{\mathbf{R}}
\newcommand{\hy}{\hat{y}}
\newcommand{\cmp}{\text{cmp}}
\newcommand{\spr}{\text{spr}}
\newcommand{\bW}{\bar{W}}
\newcommand{\RealsP}{\Reals_{+}}
\newcommand{\hW}{\hat{W}}
\newcommand{\beq}{\begin{eqnarray}}
\newcommand{\eeq}{\end{eqnarray}}

\begin{document}
%

\title{Fast Distributed Algorithms for Computing Separable Functions}

\author{Damon Mosk-Aoyama and Devavrat Shah 
%
\thanks{D. Mosk-Aoyama is with the department of 
Computer Science, Stanford University. D. Shah is
with the department of Electrical Engineering and Computer Science,
MIT.  Emails:~\{damonma@cs.stanford.edu,devavrat@mit.edu\}
}
}

\date{}

\maketitle

\begin{abstract}

The problem of computing functions of values at the nodes in a network
in a totally distributed manner, where nodes do not have unique
identities and make decisions based only on local information, has
applications in sensor, peer-to-peer, and ad-hoc networks.  The task
of computing separable functions, which can be written as linear
combinations of functions of individual variables, is studied in this
context.  Known iterative algorithms for averaging can be used to
compute the normalized values of such functions, but these algorithms
do not extend in general to the computation of the actual values of
separable functions.

The main contribution of this paper is the design of a distributed
randomized algorithm for computing separable functions.  The running
time of the algorithm is shown to depend on the running time of a
minimum computation algorithm used as a subroutine.  Using a
randomized gossip mechanism for minimum computation as the subroutine
yields a complete totally distributed algorithm for computing
separable functions.  For a class of graphs with small spectral gap,
such as grid graphs, the time used by the algorithm to compute
averages is of a smaller order than the time required by a known
iterative averaging scheme.

\end{abstract}




\section{Introduction}

The development of sensor, peer-to-peer, and ad hoc wireless networks
has stimulated interest in distributed algorithms for data
aggregation, in which nodes in a network compute a function of local
values at the individual nodes.  These networks typically do not have
centralized agents that organize the computation and communication
among the nodes.  Furthermore, the nodes in such a network may not
know the complete topology of the network, and the topology may change
over time as nodes are added and other nodes fail.  In light of the
preceding considerations, distributed computation is of vital
importance in these modern networks.

We consider the problem of computing separable functions in a
distributed fashion in this paper.  A separable function can be
expressed as the sum of the values of individual functions.  Given a
network in which each node has a number, we seek a distributed
protocol for computing the value of a separable function of the
numbers at the nodes.  Each node has its own estimate of the value of
the function, which evolves as the protocol proceeds.  Our goal is to
minimize the amount of time required for all of these estimates to be
close to the actual function value.

In this work, we are interested in {\em totally distributed}
computations, in which nodes have a local view of the state of the
network.  Specifically, an individual node does not have information
about nodes in the network other than its neighbors.  To accurately
estimate the value of a separable function that depends on the numbers
at all of the nodes, each node must obtain information about the other
nodes in the network.  This is accomplished through communication
between neighbors in the network.  Over the course of the protocol,
the global state of the network effectively diffuses to each
individual node via local communication among neighbors.

More concretely, we assume that each node in the network knows only
its neighbors in the network topology, and can contact any neighbor to
initiate a communication.  On the other hand, we assume that the nodes
do not have unique identities (i.e., a node has no unique identifier
that can be attached to its messages to identify the source of the
messages).  This constraint is natural in ad-hoc and mobile networks,
where there is a lack of infrastructure (such as IP addresses or
static GPS locations), and it limits the ability of a distributed
algorithm to recreate the topology of the network at each node.  In
this sense, the constraint also provides a formal way to distinguish
distributed algorithms that are truly local from algorithms that
operate by gathering enormous amounts of global information at all the
nodes.

The absence of identifiers for nodes makes it difficult, without
global coordination, to simply transmit every node's value throughout
the network so that each node can identify the values at all the
nodes.  As such, we develop an algorithm for computing separable
functions that relies on an {\em order- and duplicate-insensitive}
statistic \cite{ngsa} of a set of numbers, the minimum.  The algorithm
is based on properties of exponential random variables, and reduces
the problem of computing the value of a separable function to the
problem of determining the minimum of a collection of numbers, one for
each node.

This reduction leads us to study the problem of
{\em information spreading} or {\em information dissemination} in a
network.  In this problem, each node starts with a message, and the
nodes must spread the messages throughout the network using local
communication so that every node eventually has every message.
Because the minimum of a collection of numbers is not affected by the
order in which the numbers appear, nor by the presence of duplicates
of an individual number, the minimum computation required by our
algorithm for computing separable functions can be performed by any
information spreading algorithm.  Our analysis of the algorithm for
computing separable functions establishes an upper bound on its
running time in terms of the running time of the information spreading
algorithm it uses as a subroutine.

In view of our goal of distributed computation, we analyze a
{\em gossip} algorithm for information spreading.  Gossip algorithms
are a useful tool for achieving fault-tolerant and scalable
distributed computations in large networks.  In a gossip algorithm,
each node repeatedly iniatiates communication with a small number of
neighbors in the network, and exchanges information with those
neighbors.

The gossip algorithm for information spreading that we study is
randomized, with the communication partner of a node at any time
determined by a simple probabilistic choice.  We provide an upper
bound on the running time of the algorithm in terms of the
{\em conductance} of a stochastic matrix that governs how nodes choose
communication partners.  By using the gossip algorithm to compute
minima in the algorithm for computing separable functions, we obtain
an algorithm for computing separable functions whose performance on
certain graphs compares favorably with that of known iterative
distributed algorithms \cite{bgps} for computing averages in a
network.

\subsection{Related work}
\label{sec:related}

In this section, we present a brief summary of related work.
Algorithms for computing the number of distinct elements in a multiset
or data stream \cite{fm, streamdistinct} can be adapted to compute
separable functions using information spreading \cite{clkb}.  We are
not aware, however, of a previous analysis of the amount of time
required for these algorithms to achieve a certain accuracy in the
estimates of the function value when the computation is totally
distributed (i.e., when nodes do not have unique identities).  These
adapted algorithms require the nodes in the network to make use of a
common hash function.  In addition, the discreteness of the counting
problem makes the resulting algorithms for computing separable
functions suitable only for functions in which the terms in the sum
are integers.  Our algorithm is simpler than these algorithms, and can
compute functions with non-integer terms.



There has been a lot of work on the distributed computation of
averages, a special case of the problem of reaching agreement or
consensus among processors via a distributed computation.  Distributed
algorithms for reaching consensus under appropriate conditions have
been known since the classical work of Tsitsiklis
\cite{tsitsiklis-thesis} and Tsitsiklis, Bertsekas, and Athans
\cite{tba} (see also the book by Bertsekas and Tsitsiklis
\cite{pardiscomp}).  Averaging algorithms compute the ratio of the sum
of the input numbers to $n$, the number of nodes in the network, and
not the exact value of the sum.  Thus, such algorithms cannot be
extended in general to compute arbitrary separable functions.  On the
other hand, an algorithm for computing separable functions can be used
to compute averages by separately computing the sum of the input
numbers, and the number of nodes in the graph (using one as the input
at each node).


Recently, Kempe, Dobra, and Gehrke showed the existence of a
randomized iterative gossip algorithm for averaging with the optimal
averaging time \cite{kempe}.  This result was restricted to complete
graphs.  The algorithm requires that the nodes begin the computation
in an asymmetric initial state in order to compute separable
functions, a requirement that may not be convenient for large networks
that do not have centralized agents for global coordination.
Furthermore, the algorithm suffers from the possibility of oscillation
throughout its execution.

In a more recent paper, Boyd, Ghosh, Prabhakar, and Shah presented a
simpler iterative gossip algorithm for averaging that addresses some
of the limitations of the Kempe et al. algorithm \cite{bgps}.
Specifically, the algorithm and analysis are applicable to arbitrary
graph topologies.  Boyd et al. showed a connection between the
averaging time of the algorithm and the mixing time (a property that
is related to the conductance, but is not the same) of an appropriate
random walk on the graph representing the network.  They also found an
optimal averaging algorithm as a solution to a semi-definite program.

For completeness, we contrast our results for the problem of averaging
with known results.  As we shall see, iterative averaging, which has
been a common approach in the previous work, is an order slower than
our algorithm for many graphs, including ring and grid graphs.  In
this sense, our algorithm is quite different than (and has advantages
in comparison with) the known averaging algorithms.

On the topic of information spreading, gossip algorithms for
disseminating a message to all nodes in a complete graph in which
communication partners are chosen uniformly at random have been
studied for some time \cite{frieze, rumor2, epidemic}.  Karp,
Schindelhauer, Shenker, and V\"{o}cking presented a
{\em push and pull} gossip algorithm, in which communicating nodes
both send and receive messages, that disseminates a message to all $n$
nodes in a graph in $O(\log n)$ time with high probability
\cite{kssv}.  In this work, we have provided an analysis of the time
required for a gossip algorithm to disseminate $n$ messages to $n$
nodes for the more general setting of arbitrary graphs and non-uniform
random choices of communication partners.  For other related results,
we refer the reader to \cite{rumor3, gossip1, gossip2}.  We take note
of the similar (independent) recent work of Ganesh, Massouli\'{e}, and
Towsley \cite{gmt}, and Berger, Borgs, Chayes, and Saberi \cite{bbcs},
on the spread of epidemics in a network.

\subsection{Organization}

The rest of the paper is organized as follows.  Section
\ref{sec:prelim} presents the distributed computation problems we
study and an overview of our results.  In Section \ref{sec:comp}, we
develop and analyze an algorithm for computing separable functions in
a distributed manner.  Section \ref{sec:infdis} contains an analysis
of a simple randomized gossip algorithm for information spreading,
which can be used as a subroutine in the algorithm for computing
separable functions.  In Section \ref{sec:appl}, we discuss
applications of our results to particular types of graphs, and compare
our results to previous results for computing averages.  Finally, we
present conclusions and future directions in Section \ref{sec:conc}.

\section{Preliminaries and Results}
\label{sec:prelim}

We consider an arbitrary connected network, represented by an
undirected graph $G = (V, E)$, with $|V| = n$ nodes.  For notational
purposes, we assume that the nodes in $V$ are numbered arbitrarily so
that $V = \{1, \dots, n\}$.  A node, however, does not have a unique
identity that can be used in a computation.  Two nodes $i$ and $j$ can
communicate with each other if (and only if) $(i, j) \in E$.

To capture some of the resource constraints in the networks in which
we are interested, we impose a {\em transmitter gossip} constraint on
node communication.  Each node is allowed to contact at most one other
node at a given time for communication.  However, a node can be
contacted by multiple nodes simultaneously.

Let $2^{V}$ denote the power set of the vertex set $V$ (the set of all
subsets of $V$).  For an $n$-dimensional vector
$\vec{x} \in \Reals^{n}$, let $x_{1}, \dots, x_{n}$ be the components
of $\vec{x}$.
\begin{definition}
We say that a function $f : \Reals^{n} \times 2^{V} \to \Reals$
is {\em separable} if there exist functions $f_{1}, \dots, f_{n}$ such
that, for all $S \subseteq V$,
\begin{equation}
f(\vec{x}, S) = \sum_{i \in S} f_{i}(x_{i}).
\label{sepsum}
\end{equation}
\label{sepfunc}
\end{definition}

\noindent
{\bf Goal.}  Let $\cal{F}$ be the class of separable functions $f$ for
which $f_{i}(x) \geq 1$ for all $x \in \Reals$ and $i = 1, \dots, n$.
Given a function $f \in \cal{F}$, and a vector $\vec{x}$ containing
initial values $x_{i}$ for all the nodes, the nodes in the network are
to compute the value $f(\vec{x}, V)$ by a distributed computation,
using repeated communication between nodes.

\begin{note}
Consider a function $g$ for which there exist functions
$g_{1}, \dots, g_{n}$ satisfying, for all $S \subseteq V$, the
condition $g(\vec{x}, S) = \prod_{i \in S} g_{i}(x_{i})$ in lieu of
(\ref{sepsum}).  Then, $g$ is {\em logarithmic separable}, i.e.,
$f = \log_b g$ is separable.  Our algorithm for computing separable
functions can be used to compute the function $f = \log_{b} g$.  The
condition $f_{i}(x) \geq 1$ corresponds to $g_{i}(x) \geq b$ in this
case.  This lower bound of $1$ on $f_{i}(x)$ is arbitrary, although
our algorithm does require the terms $f_{i}(x_{i})$ in the sum to be
positive.
\end{note}

Before proceeding further, we list some practical situations where the
distributed computation of separable functions arises naturally.  By
definition, the sum of a set of numbers is a separable function.
\renewcommand{\labelenumi}{(\arabic{enumi})}
\begin{enumerate}
\item
{\em Summation.} Let the value at each node be $x_{i} = 1$.  Then, the
sum of the values is the number of nodes in the network.

\item
{\em Averaging.} According to Definition \ref{sepfunc}, the average of
a set of numbers is not a separable function.  However, the nodes can
estimate the separable function $\sum_{i = 1}^{n} x_{i}$ and $n$
separately, and use the ratio between these two estimates as an
estimate of the mean of the numbers.

Suppose the values at the nodes are measurements of a quantity of
interest.  Then, the average provides an unbiased maximum likelihood
estimate of the measured quantity.  For example, if the nodes are
temperature sensors, then the average of the sensed values at the
nodes gives a good estimate of the ambient temperature.
\end{enumerate}

For more sophisticated applications of a distributed averaging
algorithm, we refer the reader to \cite{distr_eigvec} and
\cite{MSZ}. 
Averaging is used for the distributed computation of the top $k$
eigenvectors of a graph in \cite{distr_eigvec}, while in \cite{MSZ}
averaging is used in a throughput-optimal distributed scheduling
algorithm in a wireless network.

\noindent{\bf Time model.} In a distributed computation, a time model
determines when nodes communicate with each other.  We consider two
time models, one synchronous and the other asynchronous, in this
paper.  The two models are described as follows.
\begin{enumerate}
\item
{\em Synchronous time model:} Time is slotted commonly across all
nodes in the network.  In any time slot, each node may contact one of
its neighbors according to a random choice that is independent of the
choices made by the other nodes.  The simultaneous communication
between the nodes satisfies the transmitter gossip constraint.

\item
{\em Asynchronous time model:} Each node has a clock that ticks at the
times of a rate $1$ Poisson process.  Equivalently, a common clock
ticks according to a rate $n$ Poisson process at times
$C_{k}, k \geq 1$, where $\{C_{k + 1} - C_{k}\}$ are i.i.d.
exponential random variables of rate $n$.  On clock tick $k$, one of
the $n$ nodes, say $I_{k}$, is chosen uniformly at random.  We
consider this global clock tick to be a tick of the clock at node
$I_{k}$.  When a node's clock ticks, it contacts one of its neighbors
at random.  In this model, time is discretized according to clock
ticks.  On average, there are $n$ clock ticks per one unit of absolute
time.
\end{enumerate}

In this paper, we measure the running times of algorithms in absolute
time, which is the number of time slots in the synchronous model, and
is (on average) the number of clock ticks divided by $n$ in the
asynchronous model.  To obtain a precise relationship between clock
ticks and absolute time in the asynchronous model, we appeal to tail
bounds on the probability that the sample mean of i.i.d. exponential
random variables is far from its expected value.  In particular, we
make use of the following lemma, which also plays a role in the
analysis of the accuracy of our algorithm for computing separable
functions.
\begin{lemma}
\label{discrete-to-cont}
For any $k \geq 1$, let $Y_{1}, \dots, Y_{k}$ be i.i.d. exponential
random variables with rate $\lambda$.  Let
$R_{k} = \frac{1}{k} \sum_{i = 1}^{k} Y_{i}$.  Then, for any
$\epsilon \in (0, 1/2)$,
\begin{eqnarray}
\Pr \left(\left|R_k - \frac{1}{\lambda}\right|
\geq \frac{\epsilon}{\lambda} \right)
& \leq & 2 \exp\left(-\frac{\epsilon^{2} k}{3}\right).
\label{e:dtoc1}
\end{eqnarray}
\end{lemma}
\begin{proof}
By definition,
$E[R_{k}] = \frac{1}{k}\sum_{i = 1}^{k} \lambda^{-1} = \lambda^{-1}$.
The inequality in (\ref{e:dtoc1}) follows directly from Cram\'{e}r's
Theorem (see \cite{dembo}, pp. $30$, $35$) and properties of
exponential random variables.
\end{proof}

A direct implication of Lemma \ref{discrete-to-cont} is the following
corollary, which bounds the probability that the absolute time $C_{k}$
at which clock tick $k$ occurs is far from its expected value.
\begin{corollary}\label{discrete-to-contc}
For $k \geq 1$, $E[C_{k}] = k/n$.  Further, for any
$\epsilon \in (0, 1/2)$,
\begin{eqnarray}
\Pr \left( \left|C_{k} - \frac{k}{n}\right|
\geq \frac{\epsilon k}{n} \right)
& \leq & 2 \exp\left(-\frac{\epsilon^{2} k}{3}\right).
\label{e:dtoc}
\end{eqnarray}
\end{corollary}

Our algorithm for computing separable functions is randomized, and is
not guaranteed to compute the exact quantity
$f(\vec{x}, V) = \sum_{i = 1}^{n} f_{i}(x_{i})$ at each node in the
network.  To study the accuracy of the algorithm's estimates, we
analyze the probability that the estimate of $f(\vec{x}, V)$ at every
node is within a $(1 \pm \epsilon)$ multiplicative factor of the true
value $f(\vec{x}, V)$ after the algorithm has run for some period of
time.  In this sense, the error in the estimates of the algorithm is
relative to the magnitude of $f(\vec{x}, V)$.

To measure the amount of time required for an algorithm's estimates to
achieve a specified accuracy with a specified probability, we define
the following quantity.  For an algorithm ${\cal C}$ that estimates
$f(\vec{x}, V)$, let $\hy_i(t)$ be the estimate of $f(\vec{x}, V)$ at
node $i$ at time $t$.  Furthermore, for notational convenience, given
$\epsilon > 0$, let $A_{i}^{\epsilon}(t)$ be the following event.
\[
A_{i}^{\epsilon}(t)
= \left\{\hy_{i}(t) \not\in
\left[(1 - \epsilon)f(\vec{x}, V),
(1 + \epsilon)f(\vec{x}, V) \right] \right\}
\]
\begin{definition}
For any $\epsilon > 0$ and $\delta \in (0, 1)$, the
($\epsilon$, $\delta$)-computing time of $\cal{C}$, denoted
$T_{\cal{C}}^{\cmp}(\epsilon, \delta)$, is
\[
T_{\cal{C}}^{\cmp}(\epsilon, \delta)
= \sup_{f \in \cal{F}} \sup_{\vec{x} \in \Reals^{n}}
\inf \Big\{\tau : \forall t \geq \tau,
\Pr \big(\cup_{i = 1}^{n} A_{i}^{\epsilon}(t) \big)
\leq \delta \Big\}.
\]
\end{definition}

\noindent
Intuitively, the significance of this definition of the
$(\epsilon, \delta)$-computing time of an algorithm $\cal{C}$ is that,
if $\cal{C}$ runs for an amount of time that is at least
$T_{\cal{C}}^{\cmp}(\epsilon, \delta)$, then the probability that the
estimates of $f(\vec{x}, V)$ at the nodes are all within a
$(1 \pm \epsilon)$ factor of the actual value of the function is at
least $1 - \delta$.

As noted before, our algorithm for computing separable functions is
based on a reduction to the problem of information spreading, which is
described as follows.  Suppose that, for $i = 1, \dots, n$, node $i$
has the one message $m_{i}$.  The task of information spreading is to
disseminate all $n$ messages to all $n$ nodes via a sequence of local
communications between neighbors in the graph.  In any single
communication between two nodes, each node can transmit to its
communication partner any of the messages that it currently holds.  We
assume that the data transmitted in a communication must be a set of
messages, and therefore cannot be arbitrary information.

Consider an information spreading algorithm $\cal{D}$, which specifies
how nodes communicate.  For each node $i \in V$, let $S_{i}(t)$ denote
the set of nodes that have the message $m_{i}$ at time $t$.  While
nodes can gain messages during communication, we assume that they do
not lose messages, so that $S_{i}(t_{1}) \subseteq S_{i}(t_{2})$ if
$t_{1} \leq t_{2}$.  Analogous to the $(\epsilon, \delta)$-computing
time, we define a quantity that measures the amount of time required
for an information spreading algorithm to disseminate all the messages
$m_{i}$ to all the nodes in the network.
\begin{definition}
For $\delta \in (0, 1)$, the $\delta$-information-spreading time
of the algorithm $\cal{D}$, denoted $T_{\cal{D}}^{\spr}(\delta)$, is
\[
T_{\cal{D}}^{\spr}(\delta)
= \inf
\left\{t : \Pr \left(\cup_{i = 1}^{n} \{S_{i}(t) \neq V\} \right)
\leq \delta \right\}.
\]
\end{definition}

In our analysis of the gossip algorithm for information spreading, we
assume that when two nodes communicate, each node can send all of its
messages to the other in a single communication.  This rather
unrealistic assumption of {\em infinite} link capacity is merely for
convenience, as it provides a simpler analytical characterization of
$T_{\cal{C}}^{\cmp}(\epsilon, \delta)$ in terms of
$T_{\cal{D}}^{\spr}(\delta)$.  Our algorithm for computing separable
functions requires only links of unit capacity.

\subsection{Our contribution}
\label{ssec:contrib}

The main contribution of this paper is the design of a distributed
algorithm to compute separable functions of node values in an
arbitrary connected network.  Our algorithm is randomized, and in
particular uses exponential random variables.  This usage of
exponential random variables is analogous to that in an 
algorithm by Cohen\footnote{We thank Dahlia Malkhi for pointing
this reference out to us.}
 for estimating the sizes of sets in a graph \cite{cohen}.  The
basis for our algorithm is the following property of the exponential
distribution.
\begin{property}
\label{p1}
Let $W_{1}, \dots, W_{n}$ be $n$ independent random variables such
that, for $i = 1, \dots, n$, the distribution of $W_{i}$ is
exponential with rate $\lambda_{i}$.  Let $\bW$ be the minimum of
$W_{1}, \dots, W_{n}$.  Then, $\bW$ is distributed as an exponential
random variable of rate $\lambda = \sum_{i = 1}^{n} \lambda_{i}$.
\end{property}
\begin{proof}
For an exponential random variable $W$ with rate $\lambda$, for any
$z \in \RealsP$,
\[
\Pr(W > z) = \exp(-\lambda z).
\]
Using this fact and the independence of the random variables $W_{i}$,
we compute $\Pr(\bW > z)$ for any $z \in \RealsP$.
\begin{eqnarray*}
\Pr(\bW > z)
& = & \Pr \left(\cap_{i = 1}^{n} \{W_{i} > z\} \right) \\
& = & \prod_{i = 1}^{n} \Pr(W_{i} > z) \\
& = & \prod_{i = 1}^{n} \exp(-\lambda_{i} z) \\
& = & \exp\left(-z \sum_{i = 1}^{n} \lambda_{i} \right).
\end{eqnarray*}
This establishes the property stated above.
\end{proof}

Our algorithm uses an information spreading algorithm as a subroutine,
and as a result its running time is a function of the running time of
the information spreading algorithm it uses.  The faster the
information spreading algorithm is, the better our algorithm performs.
Specifically, the following result provides an upper bound on the
($\epsilon$, $\delta$)-computing time of the algorithm.
\begin{theorem}
\label{thm:main1}
Given an information spreading algorithm $\cal{D}$ with
$\delta$-spreading time $T_{\cal{D}}^{\spr}(\delta)$ for
$\delta \in (0, 1)$, there exists an algorithm ${\cal{A}}$ for
computing separable functions $f \in \cal{F}$ such that, for any
$\epsilon \in (0, 1)$ and $\delta \in (0, 1)$,
\[
T_{\cal{A}}^{\cmp}(\epsilon, \delta)
= O\left( \epsilon^{-2} (1 + \log \delta^{-1})
T_{\cal{D}}^{\spr}(\delta/2) \right).
\]
\end{theorem}

Motivated by our interest in decentralized algorithms, we analyze a
simple randomized gossip algorithm for information spreading.  When
node $i$ initiates a communication, it contacts each node $j \neq i$
with probability $P_{ij}$.  With probability $P_{ii}$, it does not
contact another node.  The $n \times n$ matrix $P = [P_{ij}]$
characterizes the algorithm; each matrix $P$ gives rise to an
information spreading algorithm $\cal{P}$.  We assume that $P$ is
stochastic, and that $P_{ij} = 0$ if $i \neq j$ and $(i, j) \notin E$,
as nodes that are not neighbors in the graph cannot communicate with
each other.  Section \ref{sec:infdis} describes the data transmitted
between two nodes when they communicate.

We obtain an upper bound on the $\delta$-information-spreading time of
this gossip algorithm in terms of the {\em conductance} of the matrix
$P$, which is defined as follows.
\begin{definition}
For a stochastic matrix $P$, the conductance of $P$, denoted
$\Phi(P)$, is
\[
\Phi(P)
= \min_{S \subset V, \; 0 < |S| \leq n/2}
\frac{\sum_{ i \in S, j \notin S} P_{ij}}{ |S|}.
\]
\end{definition}

\noindent
In general, the above definition of conductance is not the same as the
classical definition \cite{sinclair}.  However, we restrict our
attention in this paper to doubly stochastic matrices $P$.  When $P$
is doubly stochastic, these two definitions are equivalent.  Note that
the definition of conductance implies that $\Phi(P) \leq 1$.
\begin{theorem}
\label{thm:main2}
Consider any doubly stochastic matrix $P$ such that if $i \neq j$ and
$(i, j) \notin E$, then $P_{ij} = 0$.  There exists an information
dissemination algorithm $\cal{P}$ such that, for any
$\delta \in (0, 1)$,
\[
T_{\cal{P}}^{\spr}(\delta)
= O\left(\frac{\log n + \log \delta^{-1}}{\Phi(P)}\right).
\]
\end{theorem}
\begin{note}
The results of Theorems \ref{thm:main1} and \ref{thm:main2} hold for
both the synchronous and asynchronous time models.  Recall that the
quantities $T_{\cal{C}}^{\cmp}(\epsilon, \delta)$ and
$T_{\cal{D}}^{\spr}(\delta)$ are defined with respect to absolute time
in both models.
\end{note}


\noindent
{\bf A comparison.} Theorems \ref{thm:main1} and \ref{thm:main2} imply
that, given a doubly stochastic matrix $P$, the time required for our
algorithm to obtain a $(1 \pm \epsilon)$ approximation with
probability at least $1 - \delta$ is
$O\lf(\frac{\epsilon^{-2} (1 + \log \delta^{-1})
(\log n + \log \delta^{-1})}{\Phi(P)}\rf)$.
When the network size $n$ and the accuracy parameters $\epsilon$ and
$\delta$ are fixed, the running time scales in proportion to
$1/\Phi(P)$, a factor that captures the dependence of the algorithm on
the matrix $P$.  Our algorithm can be used to compute the average of a
set of numbers.  For iterative averaging algorithms such as the ones
in \cite{tsitsiklis-thesis} and \cite{bgps}, the convergence time
largely depends on the mixing time of $P$, which is lower bounded by
$\Omega(1/\Phi(P))$ (see \cite{sinclair}, for example).  Thus, our
algorithm is (up to a $\log n$ factor) no slower than the fastest
iterative algorithm based on time-invariant linear dynamics.

\section{Function Computation}
\label{sec:comp}

In this section, we describe our algorithm for computing the value
$y = f(\vec{x}, V) = \sum_{i = 1}^{n} f_{i}(x_{i})$ of the separable
function $f$, where $f_{i}(x_{i}) \geq 1$.  For simplicity of
notation, let $y_{i} = f_{i}(x_{i})$.  Given $x_{i}$, each node can
compute $y_{i}$ on its own.  Next, the nodes use the algorithm shown
in Fig. \ref{compalg}, which we refer to as COMP, to compute estimates
$\hy_{i}$ of $y = \sum_{i = 1}^{n} y_{i}$.  The quantity $r$ is a
parameter to be chosen later.
\begin{figure}[htbp]
\centering
\begin{minipage}{\textwidth}
\hrulefill

\noindent
{\bf Algorithm COMP}

\renewcommand{\labelenumi}{{\bf \arabic{enumi}.}}
\renewcommand{\labelenumii}{(\alph{enumii})}
\begin{enumerate}

\item[{\bf 0.}]
Initially, for $i = 1, \dots, n$, node $i$ has the value
$y_{i} \geq 1$.

\item
Each node $i$ generates $r$ independent random numbers
$W_{1}^{i}, \dots, W_{r}^{i}$, where the distribution of each
$W_{\ell}^{i}$ is exponential with rate $y_{i}$ (i.e., with mean
$1/y_{i}$).

\item
\label{minstep}
Each node $i$ computes, for $\ell = 1, \dots, r$, an estimate
$\hW_{\ell}^{i}$ of the minimum
$\bW_{\ell} = \min_{i = 1}^{n} W_{\ell}^{i}$.  This computation can be
done using an information spreading algorithm as described below.

\item
Each node $i$ computes
$\hy_{i} = \frac{r}{\sum_{\ell = 1}^{r} \hW_{\ell}^{i}}$ as its
estimate of $\sum_{i = 1}^{n} y_{i}$.
\end{enumerate}
\hrulefill
\end{minipage}
\caption{An algorithm for computing separable functions.}
\label{compalg}
\end{figure}

We describe how the minimum is computed as required by step
{\bf \ref{minstep}} of the algorithm in Section \ref{ssec:minima}.
The running time of the algorithm COMP depends on the running time of
the algorithm used to compute the minimum.

Now, we show that COMP effectively estimates the function value $y$
when the estimates $\hW_{\ell}^{i}$ are all correct by providing a
lower bound on the conditional probability that the estimates produced
by COMP are all within a $(1 \pm \epsilon)$ factor of $y$.
\begin{lemma}
\label{lem:estaccuracy}
Let $y_{1}, \dots, y_{n}$ be real numbers (with $y_{i} \geq 1$ for
$i = 1, \dots, n$), $y = \sum_{i = 1}^{n} y_{i}$, and
$\bW = (\bW_{1}, \dots, \bW_{r})$, where the $\bW_{\ell}$ are as
defined in the algorithm COMP.  For any node $i$, let
$\hW^{i} = (\hW_{1}^{i}, \dots, \hW_{r}^{i})$, and let $\hy_{i}$ be
the estimate of $y$ obtained by node $i$ in COMP.  For any
$\epsilon \in (0, 1/2)$,
\[
\begin{split}
\Pr & \left( \cup_{i = 1}^{n} \left\{
\left|\hy_{i} - y \right| > 2 \epsilon y \right\}
\mid \forall i \in V, \: \hW^{i} = \bW \right) 
 \leq 2\exp\left(-\frac{\epsilon^{2} r}{3} \right).
\end{split}
\]
\end{lemma}

\begin{proof}
Observe that the estimate $\hy_{i}$ of $y$ at node $i$ is a function
of $r$ and $\hW^{i}$.  Under the hypothesis that $\hW^{i} = \bW$ for
all nodes $i \in V$, all nodes produce the same estimate
$\hy = \hy_{i}$ of $y$.  This estimate is
$\hy = r \left(\sum_{\ell = 1}^{r} \bW_{\ell} \right)^{-1}$, and so
$\hy^{-1} = \left(\sum_{\ell = 1}^{r} \bW_{\ell} \right)r^{-1}$.

Property \ref{p1} implies that each of the $n$ random variables
$\bW_{1}, \dots, \bW_{r}$ has an exponential distribution with rate
$y$.  From Lemma \ref{discrete-to-cont}, it follows that for any
$\epsilon \in (0, 1/2)$,
\begin{equation}
\begin{split}
\Pr & \left( \left|\hy^{-1} - \frac{1}{y}\right|
> \frac{\epsilon}{y}
\;\Big|\; \forall i \in V, \: \hW^{i} = \bW \right) 
~ \leq 2\exp\left(-\frac{\epsilon^2 r}{3}\right).
\end{split}
\label{e1}
\end{equation}
This inequality bounds the conditional probability of the event
$\{\hy^{-1} \not\in
[(1 - \epsilon) y^{-1}, (1 + \epsilon)y^{-1}]\}$,
which is equivalent to the event
$\{\hy \not\in [(1 + \epsilon)^{-1}y, (1 - \epsilon)^{-1}y]\}$.
Now, for $\epsilon \in (0, 1/2)$,
\begin{equation}
\begin{split}
(1 - \epsilon)^{-1} & \in
\left[ 1 + \epsilon, 1 + 2\epsilon \right], 
~ (1 + \epsilon)^{-1}
~ \in \left[1 - \epsilon, 1 - 2\epsilon/3\right].
\end{split}
\label{e2}
\end{equation}
Applying the inequalities in (\ref{e1}) and (\ref{e2}), we conclude
that for $\epsilon \in (0, 1/2)$,
\[
\begin{split}
\Pr & \left(\left| \hy - y \right| > 2 \epsilon y
\mid \forall i \in V, \: \hW^{i} = \bW \right) ~
 \leq 2 \exp\left(-\frac{\epsilon^2 r}{3}\right).
\end{split}
\]

\noindent
Noting that the event
$\cup_{i = 1}^{n} \{|\hy_{i} - y| > 2 \epsilon y\}$ is equivalent to
the event $\{|\hy - y| > 2 \epsilon y\}$ when $\hW^{i} = \bW$ for all
nodes $i$ completes the proof of Lemma \ref{lem:estaccuracy}.
\end{proof}

\subsection{Using information spreading to compute minima}
\label{ssec:minima}

We now elaborate on step {\bf \ref{minstep}} of the algorithm COMP.
Each node $i$ in the graph starts this step with a vector
$W^{i} = (W_{1}^{i}, \dots, W_{r}^{i})$, and the nodes seek the vector
$\bW = (\bW_{1}, \dots, \bW_{r})$, where
$\bW_{\ell} = \min_{i = 1}^{n} W_{\ell}^{i}$.  In the information
spreading problem, each node $i$ has a message $m_{i}$, and the nodes
are to transmit messages across the links until every node has every
message.

If all link capacities are infinite (i.e., in one time unit, a node
can send an arbitrary amount of information to another node), then an
information spreading algorithm $\cal{D}$ can be used directly to
compute the minimum vector $\bW$.  To see this, let the message
$m_{i}$ at the node $i$ be the vector $W^{i}$, and then apply the
information spreading algorithm to disseminate the vectors.  Once
every node has every message (vector), each node can compute $\bW$ as
the component-wise minimum of all the vectors.  This implies that the
running time of the resulting algorithm for computing $\bW$ is the
same as that of the information spreading algorithm.

The assumption of infinite link capacities allows a node to transmit
an arbitrary number of vectors $W^{i}$ to a neighbor in one time unit.
A simple modification to the information spreading algorithm, however,
yields an algorithm for computing the minimum vector $\bW$ using links
of capacity $r$.  To this end, each node $i$ maintains a single
$r$-dimensional vector $w^{i}(t)$ that evolves in time, starting with
$w^{i}(0) = W^{i}$.

Suppose that, in the information dissemination algorithm, node $j$
transmits the messages (vectors) $W^{i_{1}}, \dots, W^{i_{c}}$ to node
$i$ at time $t$.  Then, in the minimum computation algorithm, $j$
sends to $i$ the $r$ quantities $w_{1}, \dots, w_{r}$, where
$w_{\ell} = \min_{u = 1}^{c} W_{\ell}^{i_{u}}$.  The node $i$ sets
$w_{\ell}^{i}(t^{+}) = \min(w_{\ell}^{i}(t^{-}), w_{\ell})$ for
$\ell = 1, \dots, r$, where $t^{-}$ and $t^{+}$ denote the times
immediately before and after, respectively, the communication.  At any
time $t$, we will have $w^{i}(t) = \bW$ for all nodes $i \in V$ if, in
the information spreading algorithm, every node $i$ has all the
vectors $W^{1}, \dots, W^{n}$ at the same time $t$.  In this way, we
obtain an algorithm for computing the minimum vector $\bW$ that uses
links of capacity $r$ and runs in the same amount of time as the
information spreading algorithm.

An alternative to using links of capacity $r$ in the computation of
$\bW$ is to make the time slot $r$ times larger, and impose a unit
capacity on all the links.  Now, a node transmits the numbers
$w_{1}, \dots, w_{r}$ to its communication partner over a period of
$r$ time slots, and as a result the running time of the algorithm for
computing $\bW$ becomes greater than the running time of the
information spreading algorithm by a factor of $r$.  The preceding
discussion, combined with the fact that nodes only gain messages as an
information spreading algorithm executes, leads to the following
lemma.
\begin{lemma}
\label{lem:mininfdis}
Suppose that the COMP algorithm is implemented using an information
spreading algorithm $\cal{D}$ as described above.  Let $\hW^{i}(t)$
denote the estimate of $\bW$ at node $i$ at time $t$.  For any
$\delta \in (0, 1)$, let $t_{m} = r T_{\cal{D}}^{\spr}(\delta)$.
Then, for any time $t \geq t_{m}$, with probability at least
$1 - \delta$, $\hW^{i}(t) = \bW$ for all nodes $i \in V$.
\end{lemma}

Note that the amount of data communicated between nodes during the
algorithm COMP depends on the values of the exponential random
variables generated by the nodes.  Since the nodes compute minima of
these variables, we are interested in a probabilistic lower bound on
the values of these variables (for example, suppose that the nodes
transmit the values $1/W_{\ell}^{i}$ when computing the minimum
$\bW_{\ell} = 1/\max_{i = 1}^{n} \{1/W_{\ell}^{i}\}$).
To this end, we use the fact that each $\bW_{\ell}$ is an exponential
random variable with rate $y$ to obtain that, for any constant
$c > 1$, the probability that any of the minimum values $\bW_{\ell}$
is less than $1/B$ (i.e., any of the inverse values $1/W_{\ell}^{i}$
is greater than $B$) is at most $\delta/c$, where $B$ is proportional
to $cry/\delta$.

\subsection{Proof of Theorem \ref{thm:main1}}

Now, we are ready to prove Theorem \ref{thm:main1}.  In particular, we
will show that the COMP algorithm has the properties claimed in
Theorem \ref{thm:main1}. To this end, consider using an information spreading algorithm $\cal{D}$ with
$\delta$-spreading time $T_{\cal{D}}^{\spr}(\delta)$ for
$\delta \in (0, 1)$ as the subroutine in the COMP algorithm.  For any
$\delta \in (0, 1)$, let $\tau_{m} = rT_{\cal{D}}^{\spr}(\delta/2)$.
By Lemma \ref{lem:mininfdis}, for any time $t \geq \tau_{m}$, the
probability that $\hW^{i} \neq \bW$ for any node $i$ at time $t$ is at
most $\delta/2$.

On the other hand, suppose that $\hW^{i} = \bW$ for all nodes $i$ at
time $t \geq \tau_{m}$.  For any $\epsilon \in (0, 1)$, by choosing
$r \geq 12 \epsilon^{-2} \log (4 \delta^{-1})$ so that
$r = \Theta(\epsilon^{-2}(1 + \log \delta^{-1}))$, we obtain from
Lemma \ref{lem:estaccuracy} that
\begin{equation}
\begin{split}
\Pr & \left(\cup_{i = 1}^{n} \left\{\hat{y}_{i}
\not\in \left[ (1 - \epsilon) y, (1 + \epsilon) y \right] \right\}
\mid \forall i \in V, \: \hW^{i} = \bW \right) ~
~ \leq \delta/2.
\end{split}
\label{e3}
\end{equation}
Recall that $T_{COMP}^{\cmp}(\epsilon, \delta)$ is the smallest time
$\tau$ such that, under the algorithm COMP, at any time $t \geq \tau$,
all the nodes have an estimate of the function value $y$ within a
multiplicative factor of $(1 \pm \epsilon)$ with probability at least
$1 - \delta$.  By a straightforward union bound of events and
(\ref{e3}), we conclude that, for any time $t \geq \tau_{m}$,
\[
\Pr \left(\cup_{i = 1}^{n} \left\{\hat{y}_{i} \not\in
\left[ (1 - \epsilon) y, (1 + \epsilon) y \right] \right\} \right)
\leq \delta.
\]
For any $\epsilon \in (0, 1)$ and $\delta \in (0, 1)$, we now have, by
the definition of $(\epsilon, \delta)$-computing time,
\begin{eqnarray*}
T_{COMP}^{\cmp}(\epsilon, \delta)
& \leq & \tau_{m} \\
& = & O \left(\epsilon^{-2} (1 + \log \delta^{-1})
T_{\cal{D}}^{\spr}(\delta/2) \right).
\end{eqnarray*}
This completes the proof of Theorem \ref{thm:main1}.

\section{Information spreading}
\label{sec:infdis}

In this section, we analyze a randomized gossip algorithm for
information spreading.  The method by which nodes choose partners to
contact when initiating a communication and the data transmitted
during the communication are the same for both time models defined in
Section \ref{sec:prelim}.  These models differ in the times at which
nodes contact each other: in the asynchronous model, only one node can
start a communication at any time, while in the synchronous model all
the nodes can communicate in each time slot.


The information spreading algorithm that we study is presented in
Fig. \ref{spreadalg}, which makes use of the following notation.  Let
$M_{i}(t)$ denote the set of messages node $i$ has at time $t$.
Initially, $M_{i}(0) = \{m_{i}\}$ for all $i \in V$.  For a
communication that occurs at time $t$, let $t^{-}$ and $t^{+}$
denote the times immediately before and after, respectively, the
communication occurs.

As mentioned in Section \ref{ssec:contrib}, the nodes choose
communication partners according to the probability distribution
defined by an $n \times n$ matrix $P$.  The matrix $P$ is
non-negative and stochastic, and satisfies $P_{ij} = 0$ for any pair
of nodes $i \neq j$ such that $(i, j) \not\in E$.  For each such
matrix $P$, there is an instance of the information spreading
algorithm, which we refer to as SPREAD($P$).
\begin{figure}[htbp]
\centering
\begin{minipage}{\textwidth}
\hrulefill

\noindent
{\bf Algorithm SPREAD($P$)}

\noindent
When a node $i$ initiates a communication at time $t$:

\renewcommand{\labelenumi}{{\bf \arabic{enumi}.}}
\begin{enumerate}

\item
Node $i$ chooses a node $u$ at random, and contacts $u$.  The choice
of the communication partner $u$ is made independently of all other
random choices, and the probability that node $i$ chooses any node $j$
is $P_{ij}$.

\item
Nodes $u$ and $i$ exchange all of their messages, so that
\[
M_{i}(t^{+}) = M_u(t^+) =  M_{i}(t^{-}) \cup M_{u}(t^{-}).
\]
\end{enumerate}
\hrulefill
\end{minipage}
\caption{A gossip algorithm for information spreading.}
\label{spreadalg}
\end{figure}

We note that the data transmitted between two communicating nodes in
SPREAD conform to the {\em push and pull mechanism}.  That is, when node $i$
contacts node $u$ at time $t$, both nodes $u$ and $i$ exchange all of
their information with each other.  We also note that the
description in the algorithm  assumes that the communication
links in the network have infinite capacity.  As discussed in Section
\ref{ssec:minima}, however, an information spreading algorithm that
uses links of infinite capacity can be used to compute minima using
links of unit capacity.

This algorithm is simple, distributed, and satisfies the transmitter
gossip constraint.  We now present analysis of the information
spreading time of SPREAD($P$) for doubly stochastic matrices $P$ in
the two time models.  The goal of the analysis is to prove Theorem
\ref{thm:main2}.  To this end, for any $i \in V$, let
$S_{i}(t) \subseteq V$ denote the set of nodes that have the message
$m_{i}$ after any communication events that occur at absolute time $t$
(communication events occur on a global clock tick in the asynchronous
time model, and in each time slot in the synchronous time model).  At
the start of the algorithm, $S_{i}(0) = \{i\}$.

\subsection{Asynchronous model}

As described in Section \ref{sec:prelim}, in the asynchronous time
model the global clock ticks according to a Poisson process of rate
$n$, and on a tick one of the $n$ nodes is chosen uniformly at random.
This node initiates a communication, so the times at which the
communication events occur correspond to the ticks of the clock.  On
any clock tick, at most one pair of nodes can exchange messages by
communicating with each other.

Let $k \geq 0$ denote the index of a clock tick.  Initially, $k = 0$,
and the corresponding absolute time is $0$.  For simplicity of
notation, we identify the time at which a clock tick occurs with its
index, so that $S_{i}(k)$ denotes the set of nodes that have the
message $m_{i}$ at the end of clock tick $k$.  The following lemma
provides a bound on the number of clock ticks required for every node
to receive every message.
\begin{lemma}
\label{lem:asynchticks}
For any $\delta \in (0, 1)$, define 
\[
K(\delta)
= \inf\{k \geq 0:
\Pr(\cup_{i = 1}^{n} \{S_{i}(k) \neq V\}) \leq \delta \}.
\]
Then,
\[
K(\delta)
= O\left( n \frac{\log n + \log \delta^{-1}}{\Phi(P)}\right).
\]
\end{lemma}

\begin{proof}
Fix any node $v \in V$.  We study the evolution of the size of the set
$S_{v}(k)$.  For simplicity of notation, we drop the subscript $v$,
and write $S(k)$ to denote $S_{v}(k)$.

Note that $|S(k)|$ is monotonically non-decreasing over the course of
the algorithm, with the initial condition $|S(0)| = 1$.  For the
purpose of analysis, we divide the execution of the algorithm into two
phases based on the size of the set $S(k)$.  In the first phase,
$|S(k)| \leq n/2$, and in the second phase $|S(k)| > n/2$.

Under the gossip algorithm, after clock tick $k + 1$, we have either
$|S(k + 1)| = |S(k)|$ or $|S(k + 1)| = |S(k)| + 1$.  Further, the size
increases if a node $i \in S(k)$ contacts a node $j \notin S(k)$, as
in this case $i$ will push the message $m_{v}$ to $j$.  For each such
pair of nodes $i$, $j$, the probability that this occurs on clock tick
$k + 1$ is $P_{ij}/n$.  Since only one node is active on each clock
tick,
\begin{equation}
E[|S(k + 1)| - |S(k)| \mid S(k)]
\geq \sum_{i \in S(k), j \notin S(k)} \frac{P_{ij}}{n}.
\label{expinc}
\end{equation}

\noindent
When $|S(k)| \leq n/2$, it follows from (\ref{expinc}) and the
definition of the conductance $\Phi(P)$ of $P$ that
\begin{eqnarray}
E[|S(k + 1)| - |S(k)| \mid S(k)]
& \geq & \frac{|S(k)|}{n}
\frac{\sum_{i \in S(k), j \notin S(k)} P_{ij} }{|S(k)|}
\notag \\
& \geq &
\frac{|S(k)|}{n} \min_{S \subset V, \; 0 < |S| \leq n/2}
\frac{\sum_{ i \in S, j \notin S} P_{ij}}{|S|}
\notag \\
& = & \frac{|S(k)|}{n} \Phi(P)
\notag \\
& = & |S(k)| \hat{\Phi},
\label{e4}
\end{eqnarray}

\noindent
where $\hat{\Phi} = \frac{\Phi(P)}{n}$.

We seek an upper bound on the duration of the first phase.  To this
end, let
\begin{equation*}
Z(k) = \frac{\exp \left(\frac{\hat{\Phi}}{4} k \right)}{|S(k)|}.
\end{equation*}

\noindent
Define the stopping time $L = \inf \{k : |S(k)| > n/2\}$, and
$L \land k = \min(L, k)$.  If $|S(k)| > n/2$, then
$L \land (k + 1) = L \land k$, and thus
$E[Z(L \land (k + 1)) \mid S(L \land k)] = Z(L \land k)$.

Now, suppose that $|S(k)| \leq n/2$, in which case
$L \land (k + 1) = (L \land k) + 1$.  The function $g(z) = 1/z$ is
convex for $z > 0$, which implies that, for $z_{1}, z_{2} > 0$,
\begin{equation}
g(z_{2}) \geq g(z_{1}) + g'(z_{1})(z_{2} - z_{1}).
\label{convderunder}
\end{equation}

\noindent
Applying \eqref{convderunder} with $z_{1} = |S(k + 1)|$ and
$z_{2} = |S(k)|$ yields
\begin{equation*}
\frac{1}{|S(k + 1)|}
\leq \frac{1}{|S(k)|}
- \frac{1}{|S(k + 1)|^{2}} (|S(k + 1)| - |S(k)|).
\end{equation*}

\noindent
Since $|S(k + 1)| \leq |S(k)| + 1 \leq 2|S(k)|$, it follows that
\begin{equation}
\frac{1}{|S(k + 1)|}
\leq \frac{1}{|S(k)|}
- \frac{1}{4 |S(k)|^{2}} (|S(k + 1)| - |S(k)|).
\label{invsizeupper}
\end{equation}

Combining \eqref{e4} and \eqref{invsizeupper}, we obtain that, if
$|S(k)| \leq n/2$, then
\begin{equation*}
E \left[\frac{1}{|S(k + 1)|} \;\Big|\; S(k) \right]
\leq \frac{1}{|S(k)|} \left(1 - \frac{\hat{\Phi}}{4} \right)
\leq \frac{1}{|S(k)|} \exp \left(-\frac{\hat{\Phi}}{4} \right),
\end{equation*}

\noindent
as $1 - z \leq \exp(-z)$ for $z \geq 0$.  This implies that
\begin{eqnarray*}
E[Z(L \land (k + 1)) \mid S(L \land k)]
& = & E \left[
\frac{\exp \left(\frac{\hat{\Phi}}{4} (L \land (k + 1)) \right)}
{|S(L \land (k + 1))|} \;\bigg|\; S(L \land k) \right]
\notag \\
& = & \exp \left(\frac{\hat{\Phi}}{4} (L \land k) \right)
\exp \left(\frac{\hat{\Phi}}{4} \right)
E \left[ \frac{1}
{|S((L \land k) + 1)|} \;\Big|\; S(L \land k) \right]
\notag \\
& \leq & \exp \left(\frac{\hat{\Phi}}{4} (L \land k) \right)
\exp \left(\frac{\hat{\Phi}}{4} \right)
\exp \left(-\frac{\hat{\Phi}}{4} \right)
\frac{1}{|S(L \land k)|}
\notag \\
& = & Z(L \land k),
\end{eqnarray*}

\noindent
and therefore $Z(L \land k)$ is a supermartingale.

Since $Z(L \land k)$ is a supermartingale, we have the inequality
$E[Z(L \land k)] \leq E[Z(L \land 0)] = 1$ for any $k > 0$, as
$Z(L \land 0) = Z(0) = 1$.  The fact that the set $S(k)$ can contain
at most the $n$ nodes in the graph implies that
\begin{equation*}
Z(L \land k)
= \frac{\exp \left(\frac{\hat{\Phi}}{4} (L \land k) \right)}
{|S(L \land k)|}
\geq \frac{1}{n} \exp \left(\frac{\hat{\Phi}}{4} (L \land k) \right),
\end{equation*}

\noindent
and so
\begin{equation*}
E \left[\exp \left(\frac{\hat{\Phi}}{4} (L \land k) \right) \right]
\leq n E[Z(L \land k)] \leq n.
\end{equation*}

\noindent
Because $\exp(\hat{\Phi}(L \land k)/4) \uparrow \exp (\hat{\Phi}L/4)$
as $k \to \infty$, the monotone convergence theorem implies that
\begin{equation*}
E \left[\exp \left(\frac{\hat{\Phi}L}{4} \right) \right] \leq n.
\end{equation*}

\noindent
Applying Markov's inequality, we obtain that, for
$k_{1} = 4(\ln 2 + 2 \ln n + \ln (1/\delta))/\hat{\Phi}$,
\begin{eqnarray}
\Pr (L > k_{1})
& = & \Pr \left(\exp \left(\frac{\hat{\Phi} L}{4} \right)
> \frac{2n^{2}}{\delta} \right)
\notag \\
& < & \frac{\delta}{2n}.
\label{e6a}
\end{eqnarray}

For the second phase of the algorithm, when $|S(k)| > n/2$, we study
the evolution of the size of the set of nodes that do not have the
message, $|S(k)^{c}|$.  This quantity will decrease as the message
spreads from nodes in $S(k)$ to nodes in $S(k)^{c}$.  For simplicity,
let us consider restarting the process from clock tick $0$ after $L$
(i.e., when more than half the nodes in the graph have the message),
so that we have $|S(0)^{c}| \leq n/2$.

In clock tick $k + 1$, a node $j \in S(k)^{c}$ will receive the
message if it contacts a node $i \in S(k)$ and pulls the message from
$i$.  As such,
\begin{equation*}
E[|S(k)^{c}| - |S(k + 1)^{c}| \mid S(k)^{c}]
\geq \sum_{j \in S(k)^{c}, i \notin S(k)^{c}} \frac{P_{ji}}{n},
\end{equation*}

\noindent
and thus
\begin{eqnarray}
\notag
E[|S(k + 1)^{c}| \mid S(k)^{c}]
& \leq & |S(k)^c|
- \frac{\sum_{j \in S(k)^{c}, i \notin S(k)^c} P_{ji}}{n}
\notag \\
& = & |S(k)^c|
\lf(1
- \frac{\sum_{j \in S(k)^c, i \notin S(k)^c} P_{ji}}{n |S(k)^c|} \rf)
\notag \\
& \leq & |S(k)^c| \lf( 1 - \hat{\Phi} \rf).
\label{e:5}
\end{eqnarray}

We note that this inequality holds even when $|S(k)^{c}| = 0$, and as
a result it is valid for all clock ticks $k$ in the second phase.
Repeated application of \eqref{e:5} yields
\begin{eqnarray*}
E[|S(k)^{c}|]
& = & E[E[|S(k)^{c}| \mid S(k - 1)^{c}]] \\
& \leq & \left(1 - \hat{\Phi} \right)E[|S(k - 1)^{c}|] \\
& \leq & \left(1 - \hat{\Phi} \right)^{k} E[|S(0)^{c}|] \\
& \leq & \exp \left(-\hat{\Phi} k \right) \left(\frac{n}{2} \right)
\end{eqnarray*}

For
$k_{2} = \ln (n^{2}/\delta)/2\hat{\Phi} =
(2\ln n + \ln (1/\delta))/\hat{\Phi}$,
we have $E[|S(k_{2})^{c}|] \leq \delta/(2n)$.  Markov's inequality now
implies the following upper bound on the probability that not all of
the nodes have the message at the end of clock tick $k_{2}$ in the
second phase.
\begin{eqnarray}
\Pr(|S(k_{2})^{c}| > 0) & = & \Pr(|S(k_{2})^{c}| \geq 1)
\notag \\
& \leq & E[|S(k_{2})^{c}|]
\notag \\
& \leq & \frac{\delta}{2n}.
\label{e8}
\end{eqnarray}

Combining the analysis of the two phases, we obtain that, for
$k' = k_{1} + k_{2} = O((\log n + \log \delta^{-1})/\hat{\Phi})$,
$\Pr(S_{v}(k') \neq V) \leq \delta/n$.  Applying the union bound over
all the nodes in the graph, and recalling that
$\hat{\Phi} = \Phi(P)/n$, we conclude that
\begin{eqnarray*}
K(\delta)
& \leq & k' 
~ = ~ O\left(n \frac{\log n + \log \delta^{-1}}{\Phi(P)}\right).
\end{eqnarray*}
This completes the proof of Lemma \ref{lem:asynchticks}.
\end{proof}

To extend the bound in Lemma \ref{lem:asynchticks} to absolute time,
observe that Corollary \ref{discrete-to-contc} implies that the
probability that
$\kappa = K(\delta/3) + 27 \ln (3/\delta) =
O(n(\log n + \log \delta^{-1})/\Phi(P))$
clock ticks do not occur in absolute time
$(4/3) \kappa/n = O((\log n + \log \delta^{-1})/\Phi(P))$ is at most
$2 \delta/3$.  Applying the union bound now yields
$T_{SPREAD(P)}^{\spr}(\delta) =
O((\log n + \log \delta^{-1})/\Phi(P))$,
thus establishing the  upper bound in Theorem \ref{thm:main2} for the
asynchronous time model.

\subsection{Synchronous model}

In the synchronous time model, in each time slot every node contacts a
neighbor to exchange messages.  Thus, $n$ communication events may
occur simultaneously.  Recall that absolute time is measured in rounds
or time slots in the synchronous model.

The analysis of the randomized gossip algorithm for information
spreading in the synchronous model is similar to the analysis for the
asynchronous model.  However, we need additional analytical arguments
to reach analogous conclusions due to the technical challenges
presented by multiple simultaneous transmissions.

In this section, we sketch a proof of the time bound in Theorem
\ref{thm:main2},
$T_{SPREAD(P)}^{\spr}(\delta) =
O((\log n + \log \delta^{-1})/\Phi(P))$,
for the synchronous time model.  Since the proof follows a similar
structure as the proof of Lemma \ref{lem:asynchticks}, we only point
out the significant differences.

As before, we fix a node $v \in V$, and study the evolution of the
size of the set $S(t) = S_{v}(t)$.  Again, we divide the execution of
the algorithm into two phases based on the evolution of $S(t)$: in the
first phase $|S(t)| \leq n/2$, and in the second phase
$|S(t)| > n/2$. In the first phase, we analyze the increase in
$|S(t)|$, while in the second we study the decrease in $|S(t)^{c}|$.
For the purpose of analysis, in the first phase we ignore the effect
of the increase in $|S(t)|$ due to the {\em pull} aspect of protocol:
that is, when node $i$ contacts node $j$, we assume (for the purpose
of analysis) that $i$ sends the messages it has to $j$, but that $j$
does not send any messages to $i$.  Clearly, an upper bound obtained
on the time required for every node to receive every message under
this restriction is also an upper bound for the actual algorithm.

Consider a time slot $t + 1$ in the first phase.  For $j \notin S(t)$,
let $X_{j}$ be an indicator random variable that is $1$ if node $j$
receives the message $m_{v}$ via a push from some node $i \in S(t)$ in
time slot $t + 1$, and is $0$ otherwise.  The probability that $j$
does not receive $m_{v}$ via a push is the probability that no node
$i \in S(t)$ contacts $j$, and so
\begin{eqnarray}
E[X_{j} \mid S(t)]
& = & 1 - \Pr(X_{j} = 0 \mid S(t))
\notag \\
& = & 1 - \prod_{i \in S(t)} (1 - P_{ij})
\notag \\
& \geq & 1 - \prod_{i \in S(t)} \exp(-P_{ij})
\notag \\
& = & 1 - \exp \left(-\sum_{i \in S(t)} P_{ij} \right).
\label{pullproblower}
\end{eqnarray}

\noindent
The Taylor series expansion of $\exp(-z)$ about $z = 0$ implies that,
if $0 \leq z \leq 1$, then
\begin{equation}
\exp(-z) \leq 1 - z + z^{2}/2 \leq 1 - z + z/2 = 1 - z/2.
\label{taylorexpsecondterm}
\end{equation}

\noindent
For a doubly stochastic matrix $P$, we have
$0 \leq \sum_{i \in S(t)} P_{ij} \leq 1$, and so we can combine
\eqref{pullproblower} and \eqref{taylorexpsecondterm} to obtain
\begin{equation*}
E[X_{j} \mid S(t)]
\geq \frac{1}{2} \sum_{i \in S(t)} P_{ij}.
\end{equation*}

By linearity of expectation,
\begin{eqnarray*}
E[|S(t + 1)| - |S(t)| \mid S(t)]
& = & \sum_{j \not\in S(t)} E[X_{j} \mid S(t)]
\notag \\
& \geq & \frac{1}{2} \sum_{i \in S(t), j \not\in S(t)} P_{ij}
\notag \\
& = & \frac{|S(t)|}{2}
\frac{\sum_{i \in S(t), j \not\in S(t)} P_{ij}}{|S(t)|}.
\end{eqnarray*}

\noindent
When $|S(t)| \leq n/2$, we have
\begin{equation}
E[|S(t + 1)| - |S(t)| \mid S(t)]
\geq |S(t)| \frac{\Phi(P)}{2}.
\label{synchexpinccond}
\end{equation}

Inequality \eqref{synchexpinccond} is analogous to inequality
\eqref{e4} for the asynchronous time model, with $\Phi(P)/2$ in the
place of $\hat{\Phi}$.  We now proceed as in the proof of Lemma
\ref{lem:asynchticks} for the asynchronous model.  Note that
$|S(t + 1)| \leq 2 |S(t)|$ here in the synchronous model because of
the restriction in the analysis to only consider the push aspect of
the protocol in the first phase, as each node in $S(t)$ can push a
message to at most one other node in a single time slot.  Repeating
the analysis from the asynchronous model leads to the conclusion that
the first phase of the algorithm ends in
$O\lf(\frac{\log n + \log \delta^{-1}}{{\Phi(P)}}\rf)$ time with
probability at least $1 - \delta/2n$.

The analysis of the second phase is the same as that presented for the
asynchronous time model, with $\hat{\Phi}$ replaced by $\Phi$.  As a
summary, we obtain that it takes at most
$O\lf(\frac{\log n + \log \delta^{-1}}{{\Phi(P)}}\rf)$ time for the
algorithm to spread all the messages to all the nodes with probability
at least $1 - \delta$.  This completes the proof of Theorem
\ref{thm:main2} for the synchronous time model.

\section{Applications}
\label{sec:appl}

We study here the application of our preceding results to several
types of graphs.  In particular, we consider complete graphs,
constant-degree expander graphs, and grid graphs.  We use grid graphs
as an example to compare the performance of our algorithm for
computing separable functions with that of a known iterative averaging
algorithm.

For each of the three classes of graphs mentioned above, we are
interested in the $\delta$-information-spreading time
$T_{SPREAD(P)}^{\spr}(\delta)$, where $P$ is a doubly stochastic
matrix that assigns equal probability to each of the neighbors of any
node.  Specifically, the probability $P_{ij}$ that a node $i$ contacts
a node $j \neq i$ when $i$ becomes active is $1/\Delta$, where
$\Delta$ is the maximum degree of the graph, and
$P_{ii} = 1 - d_{i}/\Delta$, where $d_{i}$ is the degree of $i$.
Recall from Theorem \ref{thm:main1} that the information dissemination
algorithm SPREAD($P$) can be used as a subroutine in an algorithm for
computing separable functions, with the running time of the resulting
algorithm being a function of $T_{SPREAD(P)}^{\spr}(\delta)$.

\subsection{Complete graph}

On a complete graph, the transition matrix $P$ has $P_{ii} = 0$ for
$i = 1, \dots, n$, and $P_{ij} = 1/(n - 1)$ for $j \neq i$.  This
regular structure allows us to directly evaluate the conductance of
$P$, which is $\Phi(P) \approx 1/2$.  This implies that the
($\epsilon$, $\delta$)-computing time of the algorithm for computing
separable functions based on SPREAD($P$) is
$O(\epsilon^{-2} (1 + \log \delta^{-1})(\log n + \log \delta^{-1}))$.
Thus, for a constant $\epsilon \in (0, 1)$ and $\delta = 1/n$, the
computation time scales as $O(\log^{2} n)$.

\subsection{Expander graph}

Expander graphs have been used for numerous applications, and explicit
constructions are known for constant-degree expanders \cite{rvw}.  We
consider here an undirected graph in which the maximum degree of any
vertex, $\Delta$, is a constant.  Suppose that the edge expansion of
the graph is
\begin{equation*}
\min_{S \subset V, \; 0 < |S| \leq n/2}
\frac{|F(S, S^{c})|}{|S|} = \alpha,
\end{equation*}

\noindent
where $F(S, S^{c})$ is the set of edges in the cut $(S, S^{c})$, and
$\alpha > 0$ is a constant.  The transition matrix $P$ satisfies
$P_{ij} = 1/\Delta$ for all $i \neq j$ such that $(i, j) \in E$, from
which we obtain $\Phi(P) \geq \alpha/\Delta$.  When $\alpha$ and
$\Delta$ are constants, this leads to a similar conclusion as in the
case of the complete graph: for any constant $\epsilon \in (0, 1)$ and
$\delta = 1/n$, the computation time is $O(\log^{2} n)$.

\subsection{Grid}
\label{gridsec}

We now consider a $d$-dimensional grid graph on $n$ nodes, where
$c = n^{1/d}$ is an integer.  Each node in the grid can be represented
as a $d$-dimensional vector $a = (a_{i})$, where
$a_{i} \in \{1, \dots, c\}$ for $1 \leq i \leq d$.  There is one node
for each distinct vector of this type, and so the total number of
nodes in the graph is $c^{d} = (n^{1/d})^{d} = n$.  For any two nodes
$a$ and $b$, there is an edge $(a, b)$ in the graph if and only if,
for some $i \in \{1, \dots, d\}$, $|a_{i} - b_{i}| = 1$, and
$a_{j} = b_{j}$ for all $j \neq i$.

In \cite{isogrid}, it is shown that the isoperimetric number of this
grid graph is
\begin{equation*}
\min_{S \subset V, \; 0 < |S| \leq n/2}
\frac{|F(S, S^{c})|}{|S|}
= \Theta \left(\frac{1}{c} \right)
= \Theta \left(\frac{1}{n^{1/d}} \right).
\end{equation*}

\noindent
By the definition of the edge set, the maximum degree of a node in the
graph is $2d$.  This means that $P_{ij} = 1/(2d)$ for all $i \neq j$
such that $(i, j) \in E$, and it follows that
$\Phi(P) = \Omega \left(\frac{1}{dn^{1/d}} \right)$.  Hence, for any
$\epsilon \in (0, 1)$ and $\delta \in (0, 1)$, the
($\epsilon$, $\delta$)-computing time of the algorithm for computing
separable functions is
$O(\epsilon^{-2} (1 + \log \delta^{-1})(\log n + \log \delta^{-1})
d n^{1/d})$.

\subsection{Comparison with Iterative Averaging}

We briefly contrast the performance of our algorithm for computing
separable functions with that of the iterative averaging algorithms in
\cite{tsitsiklis-thesis} \cite{bgps}.  As noted earlier, the
dependence of the performance of our algorithm is in proportion to
$1/\Phi(P)$, which is a lower bound for the iterative algorithms based
on a stochastic matrix $P$.

In particular, when our algorithm is used to compute the average of a
set of numbers (by estimating the sum of the numbers and the number of
nodes in the graph) on a $d$-dimensional grid graph, it follows from
the analysis in Section \ref{gridsec} that the amount of time required
to ensure the estimate is within a $(1 \pm \epsilon)$ factor of the
average with probability at least $1 - \delta$ is
$O(\epsilon^{-2} (1 + \log \delta^{-1})(\log n + \log \delta^{-1})
dn^{1/d})$
for any $\epsilon \in (0, 1)$ and $\delta \in (0, 1)$.  So, for a
constant $\epsilon \in (0, 1)$ and $\delta = 1/n$, the computation
time scales as $O(dn^{1/d} \log^{2} n)$ with the size of the graph,
$n$.  The algorithm in \cite{bgps} requires $\Omega(n^{2/d} \log n)$
time for this computation.  Hence, the running time of our algorithm
is (for fixed $d$, and up to logarithmic factors) the {\em square
root} of the runnning time of the iterative algorithm!  This
relationship holds on other graphs for which the spectral gap is
proportional to the square of the conductance.

\section{Conclusions and Future Work}
\label{sec:conc}

In this paper, we presented a novel algorithm for computing separable
functions in a totally distributed manner.  The algorithm is based on
properties of exponential random variables, and the fact that the
minimum of a collection of numbers is an order- and
duplicate-insensitive statistic.

Operationally, our algorithm makes use of an information spreading
mechanism as a subroutine.  This led us to the analysis of a
randomized gossip mechanism for information spreading.  We obtained an
upper bound on the information spreading time of this algorithm in
terms of the conductance of a matrix that characterizes the algorithm.

In addition to computing separable functions, our algorithm improves
the computation time for the canonical task of averaging.  For
example, on graphs such as paths, rings, and grids, the performance of
our algorithm is of a smaller order than that of a known iterative
algorithm.

We believe that our algorithm will lead to the following totally
distributed computations: (1) an approximation algorithm for convex
minimization with linear constraints; and (2) a ``packet marking''
mechanism in the Internet.  These areas, in which summation is a key
subroutine, will be topics of our future research.

\section{Acknowledgments}

We thank Ashish Goel for a useful discussion and providing
suggestions, based on previous work \cite{Goel}, when we started this
work.

\bibliographystyle{abbrv}

\bibliography{rumorspreading}

\begin{thebibliography}{10}

\bibitem{isogrid}
M.~C. Azizo{\u{g}}lu and {\"{O}}.~E{\u{g}}ecio{\u{g}}lu.
\newblock The isoperimetric number of {$d$}-dimensional {$k$}-ary arrays.
\newblock {\em International Journal of Foundations of Computer Science},
  10(3):289--300, 1999.

\bibitem{streamdistinct}
Z.~Bar-Yossef, T.~Jayram, R.~Kumar, D.~Sivakumar, and L.~Trevisan.
\newblock Counting distinct elements in a data stream.
\newblock In {\em Proceedings of RANDOM 2002}, pages 1--10, 2002.

\bibitem{bbcs}
N.~Berger, C.~Borgs, J.~T. Chayes, and A.~Saberi.
\newblock On the spread of viruses on the internet.
\newblock In {\em Proceedings of the Sixteenth Annual ACM-SIAM Symposium on
  Discrete Algorithms}, pages 301--310, 2005.

\bibitem{pardiscomp}
D.~P. Bertsekas and J.~N. Tsitsiklis.
\newblock {\em Parallel and Distributed Computation: Numerical Methods}.
\newblock Prentice Hall, 1989.

\bibitem{bgps}
S.~Boyd, A.~Ghosh, B.~Prabhakar, and D.~Shah.
\newblock Gossip algorithms: Design, analysis and applications.
\newblock In {\em Proceedings of IEEE INFOCOM 2005}, pages 1653--1664, 2005.

\bibitem{cohen}
E.~Cohen.
\newblock Size-estimation framework with applications to transitive closure and
  reachability.
\newblock {\em Journal of Computer and System Sciences}, 55(3):441--453, 1997.

\bibitem{clkb}
J.~Considine, F.~Li, G.~Kollios, and J.~Byers.
\newblock Approximate aggregation techniques for sensor databases.
\newblock In {\em Proceedings of the 20th International Conference on Data
  Engineering}, pages 449--460, 2004.

\bibitem{dembo}
A.~Dembo and O.~Zeitouni.
\newblock {\em Large Deviations Techniques and Applications}.
\newblock Springer, second edition, 1998.

\bibitem{epidemic}
A.~Demers, D.~Greene, C.~Hauser, W.~Irish, J.~Larson, S.~Shenker, H.~Sturgis,
  D.~Swinehart, and D.~Terry.
\newblock Epidemic algorithms for replicated database maintenance.
\newblock In {\em Proceedings of the Sixth Annual ACM Symposium on Principles
  of Distributed Computing}, pages 1--12, 1987.

\bibitem{Goel}
M.~Enachescu, A.~Goel, R.~Govindan, and R.~Motwani.
\newblock Scale free aggregation in sensor networks.
\newblock In {\em International Workshop on Algorithmic Aspects of Wireless
  Sensor Networks}, 2004.

\bibitem{fm}
P.~Flajolet and G.~N. Martin.
\newblock Probabilistic counting algorithms for data base applications.
\newblock {\em Journal of Computer and System Sciences}, 31(2):182--209, 1985.

\bibitem{frieze}
A.~M. Frieze and G.~R. Grimmett.
\newblock The shortest-path problem for graphs with random arc-lengths.
\newblock {\em Discrete Applied Mathematics}, 10:57--77, 1985.

\bibitem{gmt}
A.~Ganesh, L.~Massouli\'{e}, and D.~Towsley.
\newblock The effect of network topology on the spread of epidemics.
\newblock In {\em Proceedings of IEEE INFOCOM 2005}, pages 1455--1466, 2005.

\bibitem{kssv}
R.~Karp, C.~Schindelhauer, S.~Shenker, and B.~V{\"{o}}cking.
\newblock Randomized rumor spreading.
\newblock In {\em Proceedings of the 41st Annual IEEE Symposium on Foundations
  of Computer Science}, pages 565--574, 2000.

\bibitem{kempe}
D.~Kempe, A.~Dobra, and J.~Gehrke.
\newblock Gossip-based computation of aggregate information.
\newblock In {\em Proceedings of the 44th Annual IEEE Symposium on Foundations
  of Computer Science}, pages 482--491, 2003.

\bibitem{gossip1}
D.~Kempe and J.~Kleinberg.
\newblock Protocols and impossibility results for gossip-based communication
  mechanisms.
\newblock In {\em Proceedings of the 43rd Annual IEEE Symposium on Foundations
  of Computer Science}, pages 471--480, 2002.

\bibitem{gossip2}
D.~Kempe, J.~Kleinberg, and A.~Demers.
\newblock Spatial gossip and resource location protocols.
\newblock In {\em Proceedings of the 33rd Annual ACM Symposium on Theory of
  Computing}, pages 163--172, 2001.

\bibitem{distr_eigvec}
D.~Kempe and F.~McSherry.
\newblock A decentralized algorithm for spectral analaysis.
\newblock In {\em Proceedings of the 36th Annual ACM Symposium on Theory of
  Computing}, pages 561--568, 2004.

\bibitem{MSZ}
E.~Modiano, D.~Shah, and G.~Zussman.
\newblock Maximizing throughput in wireless networks via gossip.
\newblock {\em Submitted}, 2005.

\bibitem{ngsa}
S.~Nath, P.~B. Gibbons, S.~Seshan, and Z.~R. Anderson.
\newblock Synopsis diffusion for robust aggregation in sensor networks.
\newblock In {\em Proceedings of the 2nd International Conference on Embedded
  Networked Sensor Systems}, pages 250--262, 2004.

\bibitem{rumor2}
B.~Pittel.
\newblock On spreading a rumor.
\newblock {\em SIAM Journal of Applied Mathematics}, 47(1):213--223, 1987.

\bibitem{rumor3}
R.~Ravi.
\newblock Rapid rumor ramification: Approximating the minimum broadcast time.
\newblock In {\em Proceedings of the 35th Annual IEEE Symposium on Foundations
  of Computer Science}, pages 202--213, 1994.

\bibitem{rvw}
O.~Reingold, S.~Vadhan, and A.~Wigderson.
\newblock Entropy waves, the zig-zag graph product, and new constant-degree
  expanders and extractors.
\newblock In {\em Proceedings of the 41st Annual IEEE Symposium on Foundations
  of Computer Science}, pages 3--13, 2000.

\bibitem{sinclair}
A.~Sinclair.
\newblock {\em Algorithms for Random Generation and Counting: A Markov Chain
  Approach}.
\newblock Birkh{\"{a}}user, Boston, 1993.

\bibitem{tsitsiklis-thesis}
J.~N. Tsitsiklis.
\newblock {\em Problems in Decentralized Decision Making and Computation}.
\newblock PhD thesis, Department of Electrical Engineering and Computer
  Science, Massachusetts Institute of Technology, 1984.

\bibitem{tba}
J.~N. Tsitsiklis, D.~P. Bertsekas, and M.~Athans.
\newblock Distributed asynchronous deterministic and stochastic gradient
  optimization algorithms.
\newblock {\em IEEE Transactions on Automatic Control}, 31(9):803--812, 1986.

\end{thebibliography}

\end{document}